# Do Linear Dispersions of Classical Waves Mean Dirac Cones?

Jun Mei[1,3], Ying Wu[2,3], C. T. Chan[3], and Zhao-Qing Zhang[3]

[1] *Department of Physics, South China University of Technology, Guangzhou 510641, China*

[2] *Division of Mathematical and Computer Sciences and Engineering, King Abdullah University of Science and Technology (KAUST), Thuwal 23955-6900, Saudi Arabia*

[3] *Department of Physics and William Mong Institute of Nano Science and Technology, Hong Kong University of Science and Technology, Clear Water Bay, Hong Kong*

**Abstract**

By using the $\vec{k}\cdot\vec{p}$ method, we propose a first-principles theory to study the linear dispersions in phononic and photonic crystals. The theory reveals that *only* those linear dispersions created by doubly-degenerate states can be described by a reduced Hamiltonian that can be mapped into the Dirac Hamiltonian and possess a Berry phase of $-\pi$. Triply-degenerate states can also generate Dirac-like cone dispersions, but the wavefunctions transform like a spin-1 particle and the Berry phase is zero. Our theory is capable of predicting accurately the linear slopes of Dirac/Dirac-like cones at various symmetry points in a Brilliouin zone, independent of frequency and lattice structure.

phjunmei@scut.edu.cn     ying.wu@kaust.edu.sa     phchan@ust.hk     phzzhang@ust.hk

Graphene has become the center of much attention in the past several years partly due to its intriguing transport properties, such as Klein tunneling, *Zitterbewegung*, antilocalization, abnormal quantum hall effect, etc., arising from its unique band structures where the E-*k* relation is linear at the six corners of the hexagonal Brillouin zone [1]. The existence of Dirac cones in graphene can be well understood by using a tight-binding model for carbon atoms in a honeycomb lattice [1]. Recently, Dirac cones in photonic and phononic crystals have also been found at the corners of the Brillouin zones of triangular and honeycomb lattices where two bands meet [2-9], leading to the observation of many novel wave transport properties, such as classical analogues of *Zitterbewegung* and pseudo-diffusion. It was reported that linear dispersions can also occur at the Brillouin zone center of a square lattice photonic crystal, which are induced by simultaneous zero permittivity ($\varepsilon_{eff}=0$) and permeability ($\mu_{eff}=0$), and the linear dispersions could be understood from an effective medium perspective [10]. Different from the Dirac cone near the *K* point of a triangular/honeycomb lattice, which is a result of double degeneracy, the existence of the linear dispersions near the $\Gamma$ point of a square lattice is a result of triple degeneracy, i.e., accidental degeneracy of a doubly degenerate mode and a single mode. We call such type of linear dispersions Dirac-like cones. Very recently, the Dirac-like cone has also been found in elastic waves [11] and in simple cubic lattice [12]. In the past few years, various transport properties of zero-refractive-index metamaterials have been studied near the Dirac-like point and the Dirac equation was widely assumed [13] for those studies. The Dirac cones have also been found in many other photonic or plasmonic crystals in various dimensions [14-20].

The existence of linear dispersions at some symmetry points of a Brillouin zone is much more common in classical waves than in electrons. As we will see below, some exist naturally as a consequence of lattice symmetry and some can be made to occur by tuning the microstructures of a phononic/photonic crystal. Although quite a bit of effort has been devoted to this topic recently, some fundamental questions remain unanswered. For example, the physical origin of linear dispersions of a Dirac or Dirac-like cone in classical waves is not well understood. Specifically, many authors assumed that a Dirac cone can be described by the Dirac equation with two degenerate Bloch states at the *K* point as the basis of the spinors. But this has only been shown explicitly in the nearly-free-photon approximation [2], not for a realistic phononic/photonic crystal, where Bloch states are results of multiple scatterings. For graphene, the basis functions of the spinors are the atomic wave functions of two equivalent lattice sites in the unit cell of a honeycomb lattice. However, for a Dirac point in phononic/photonic crystals, this very fundamental information is still unknown to us thus far. Also, there exists no general theory that allows us to derive the linear slopes of a Dirac cone from first principles. Finally, it is not clear whether the Dirac-like cone dispersion at the Γ point can also be described by the Dirac equation, although this has been widely assumed in the literature [13]. The answer to this question is important because it is known that the Dirac equation leads to a Berry phase which in turn gives rise to anti-localization properties as found in disordered graphene.

To answer all above questions, in this Letter, we propose a theory, which generalizes the $\vec{k} \cdot \vec{p}$ method of electrons to classical waves, to study from the first principles the origin of Dirac/Dirac-like cones in phononic/photonic crystals. The theory can accurately predicted the slopes of linear dispersions at various symmetry points, independent of frequency and lattice structure. We note that the physical properties of the Dirac-like cone dispersions have been discussed within the context of effective medium theory [10, 11]. The formulation here is more general and is valid even if effective medium theory does not apply. The reduced Hamiltonian constructed from the theory reveals that only Dirac cones can be mapped into the form of massless Dirac Hamiltonian, and the basis functions of the spinors are related to the two degenerate Bloch states at the Dirac point through some non-trivial unitary transformation. Furthermore, we show that such Dirac cones possess a Berry phase of -π and, therefore, give rise to anti-localization effects. On the other hand, the Dirac-like cones with triple degeneracy cannot be mapped into the massless Dirac Hamiltonian and carry no Berry phase, and therefore, is expected to exhibit normal localization behaviors rather than anti-localization in the presence of disorder.

We consider the following acoustic wave equation in a periodic structure:

$$\nabla \cdot \left( \frac{1}{\rho_r(\vec{r})} \nabla p \right) = -\frac{\omega^2}{c_1^2} \cdot \frac{p}{B_r(\vec{r})}, \quad (1)$$

where $p$ is the pressure, $\rho_r(\vec{r}) = \rho(\vec{r})/\rho_1$ and $B_r(\vec{r}) = B(\vec{r})/B_1$ are the relative mass density and bulk modulus, respectively, and $c_1 = \sqrt{B_1/\rho_1}$ is the speed of sound in the host. Our task is to find the origin of linear dispersions near a Dirac or Dirac-like cone located at some particular high symmetry point $\vec{k}_0$ with frequency $\omega_0$, *independent of* the origin of the degeneracy. The approach is similar to the well-known $\vec{k} \cdot \vec{p}$ method for electrons. Although the method developed below is for phononic crystals, it is also valid for 2D photonic crystals by some mappings of variables.

Assuming that all the Bloch states at the $\vec{k}_0$ point are known [21], i.e., Bloch wave functions $\psi_{n\vec{k}_0}(\vec{r}) = e^{i\vec{k}_0 \cdot \vec{r}} u_{n\vec{k}_0}(\vec{r})$ and eigenfrequencies $\omega_{n0}$, where "$n$" denotes the band index, we can write the Bloch state at a wave vector $\vec{k}$ near $\vec{k}_0$ as

$$\psi_{n\vec{k}}(\vec{r}) = u_{n\vec{k}}(\vec{r}) e^{i\vec{k} \cdot \vec{r}} = \sum_j A_{nj}(\vec{k}) e^{i(\vec{k}-\vec{k}_0)\cdot \vec{r}} \psi_{j\vec{k}_0}(\vec{r}), \quad (2)$$

where the unknown periodic functions $u_{n\vec{k}}(\vec{r})$ have been expressed as linear combinations of $u_{j\vec{k}_0}(\vec{r})$. Substituting Eq. (2) into Eq. (1), we obtain

$$\sum_j A_{nj}(\vec{k}) e^{i(\vec{k}-\vec{k}_0)\cdot \vec{r}} \left[ \frac{(\omega_{n\vec{k}}^2 - \omega_{j0}^2)\psi_{j\vec{k}_0}(\vec{r})}{c_1^2 B_r(\vec{r})} + \frac{2i(\vec{k}-\vec{k}_0) \cdot \nabla \psi_{j\vec{k}_0}(\vec{r})}{\rho_r(\vec{r})} \right.$$
$$\left. + \left( i(\vec{k}-\vec{k}_0) \cdot \nabla \frac{1}{\rho_r(\vec{r})} \right) \psi_{j\vec{k}_0}(\vec{r}) - \frac{(\vec{k}-\vec{k}_0)^2 \psi_{j\vec{k}_0}(\vec{r})}{\rho_r(\vec{r})} \right] = 0 \quad (3)$$

Utilizing the orthonormal property [22] of the basis functions $\psi_{j\vec{k}_0}(\vec{r})$, i.e., $\frac{(2\pi)^2}{\Omega} \int_{unit\ cell} \psi^*_{l\vec{k}_0}(\vec{r}) \frac{1}{B_r(\vec{r})} \psi_{j\vec{k}_0}(\vec{r}) d\vec{r} = \delta_{lj}$, where $\Omega$ is the area of a unit cell, Eq. (3) can be written as

$$\sum_j \left[ \frac{\omega_{j0}^2 - \omega_{n\vec{k}}^2}{c_1^2} \delta_{lj} - P_{lj}(\vec{k}) \right] A_{nj}(\vec{k}) = 0, \tag{4}$$

where

$$P_{lj}(\vec{k}) = (\vec{k} - \vec{k}_0) \cdot \vec{p}_{lj} - (\vec{k} - \vec{k}_0)^2 q_{lj}, \tag{5}$$

with

$$\vec{p}_{lj} = i \frac{(2\pi)^2}{\Omega} \int_{unit\ cell} \psi_{l\vec{k}_0}^*(\vec{r}) \left[ \frac{2\nabla \psi_{j\vec{k}_0}(\vec{r})}{\rho_r(\vec{r})} + \left( \nabla \frac{1}{\rho_r(\vec{r})} \right) \psi_{j\vec{k}_0}(\vec{r}) \right] d\vec{r}, \tag{6}$$

and

$$q_{lj} = \frac{(2\pi)^2}{\Omega} \int_{unit\ cell} \psi_{l\vec{k}_0}^*(\vec{r}) \frac{1}{\rho_r(\vec{r})} \psi_{j\vec{k}_0}(\vec{r}) d\vec{r}. \tag{7}$$

It is easy to show that the matrices $\vec{p}_{jl}$ and $q_{jl}$ are Hermitian, i.e., $\vec{p}_{lj}^* = \vec{p}_{jl}$ and $q_{lj}^* = q_{jl}$. Equation (4) has non-trivial solutions only when the following secular equation is satisfied:

$$\det \left| H - \frac{\omega_{n\vec{k}}^2}{c_1^2} I \right| = 0, \tag{8}$$

where the matrix elements of $H$ are given by

$$H_{lj} = \frac{\omega_{j0}^2}{c_1^2} \delta_{lj} - P_{lj}. \tag{9}$$

To obtain the dispersion relation $\omega_{n\vec{k}}(\vec{k})$ from Eq. (8), one has to incorporate all Bloch states at $\vec{k}_0$. However, for our interest of linear dispersions of a Dirac/Dirac-like cone, we only have to consider the degenerate states at the Dirac/Dirac-like point in the summation of Eq. (2). Other bands only contribute to the quadratic term $\left|\vec{k} - \vec{k}_0\right|^2$ in the dispersion relation. This greatly reduces the dimension of the matrix $H$, and makes Eq. (8) analytically solvable. The analytic solution to Eq. (8) for small $\Delta k \equiv \left|\vec{k} - \vec{k}_0\right|$ can be expressed as:

$$\frac{\Delta \omega_\beta}{\Delta k} = \gamma_\beta c_1 + O(\Delta k^2); \qquad \beta = 1, 2, \ldots, s, \tag{10}$$

where $\Delta\omega_\beta \equiv \omega_{\beta\vec{k}} - \omega_0$, $\omega_0$ and $s$ are, respectively, the frequency and number of degenerate Bloch states at the Dirac/Dirac-like point. Here we have approximated $\omega_0^2 - \omega_{\beta\vec{k}}^2$ by $-2\omega_0\Delta\omega_\beta$. The linear slopes, $\gamma_\beta$, are determined by $\vec{p}_{lj}$ only, and $q_{lj}$ contributes to the $\Delta k^2$ term. Thus, the slope of the linear dispersion is proportional to the strength of the coupling between the degenerate Bloch states $\psi_{l\vec{k}_0}$ and $\psi_{j\vec{k}_0}$, as presented in the form of $\vec{p}_{lj}$. Therefore, it is clear that the coupling of the degenerate Bloch states leads to the frequency repulsion effect, which in turn gives rise to linear dispersions of a Dirac/Dirac-like cone. It should be pointed out that the above perturbative approach is *exact* to the first order in $\Delta k$ [23]. Thus, the existence of linear dispersions requires both the degeneracy of Bloch states at some high symmetry point $\vec{k}_0$ and the corresponding $\vec{p}_{lj} \neq 0$, *independent of* whether the degeneracy is accidental or due to lattice symmetry. The matrix elements of $\vec{p}_{jl}$ and $q_{jl}$ can be easily evaluated by performing numerical integration of Eqs. (6) and (7) using the knowledge of degenerate Bloch wavefunctions at $\vec{k}_0$. There are various ways to obtain the Bloch wavefunctions for phononic/photonic crystals numerically. In this work, we adopted the COMSOL Multiphysics, a commercial package based on finite-element method.

Figure 1(a) shows the band structures of a triangular array of iron cylinders, with radii of 0.3203$a$ ($a$ is the lattice constant), embedded in water host calculated by COMSOL. There exist two points, marked by "$A$" and "$B$", at which linear dispersions are found. The mass densities of iron and water are $\rho_2 = 7670 kg/m^3$ and $\rho_1 = 1000 kg/m^3$, respectively, and the corresponding longitudinal wave velocities are $c_2 = 6010 m/s$ for iron and $c_1 = 1490 m/s$ for water. Since the longitudinal wave velocity contrast between iron and water is large, the shear modes inside the iron cylinders are not important and can be ignored [24]. This has simplified our calculations.

Point *A* shows a Dirac-like cone created by accidental degeneracy of three Bloch states at the Γ point ($\vec{k}_0 = 0$), where the frequencies for the single Bloch state, $\psi_{1\Gamma}$, and the doubly-degenerate Bloch states, $\psi_{2\Gamma}$ and $\psi_{3\Gamma}$, are tuned deliberately to coincide by adjusting the radii of the iron cylinders. Figures 1(b) and 1(c) illustrate the band structures near the Γ point for smaller and larger iron cylinders, i.e., *r=0.26a* and *0.38a*, respectively. Both figures show clearly the separated single mode and doubly-degenerate modes. The three degenerate Bloch states $\psi_{1\Gamma}$, $\psi_{2\Gamma}$ and $\psi_{3\Gamma}$ at point *A* when *r* is tuned to *0.3203a* are shown in Fig. 2(a), (b) and (c), respectively. To the linear order in $k$, the reduced Hamiltonian near the point *A* has the following form:

$$H = \begin{pmatrix} 0 & i\vec{k}\cdot\vec{L}_{12} & i\vec{k}\cdot\vec{L}_{13} \\ -i\vec{k}\cdot\vec{L}_{12} & 0 & 0 \\ -i\vec{k}\cdot\vec{L}_{13} & 0 & 0 \end{pmatrix}, \tag{11}$$

where $\vec{L}_{lj} = -i\vec{p}_{lj}$, and $\vec{L}_{12}$ and $\vec{L}_{13}$ are two real vectors having the same length, $s/a$, and perpendicular to each other as required by the isotropy of dispersion, i.e., $|\vec{L}_{12}| = |\vec{L}_{13}| = s/a$ and $\vec{L}_{12} \cdot \vec{L}_{13} = 0$. By using Eq. (8), we find $\Delta\omega/k = \pm sc_1^2/(2\omega_0 a)$ and 0. From numerical simulations, we obtain $s = 10.241$. With the frequency at point *A*, $\omega_0 a/(2\pi c_1) = 1.06$, as shown in Fig. 1(a), we find $\Delta\omega/k = \pm 0.769 c_1$ and 0. Obviously, the first two correspond to the two linear dispersions and the third one corresponds to the flat band shown near the point *A*. These results agree excellently with the band structure calculations as can be seen from Fig. 2(d). It is worth mentioning that if wave energy is localized within the scatterers, then the slopes of the Dirac-like cone can also be calculated by using a tight-binding model proposed in Ref. [12]. However, our theory does not have such restrictions as shown in Fig. 2, in which the wave energy is mainly concentrated in the host.

From Eq. (11), we find the following Bloch states near point *A*:
$$\Psi_+^T(\vec{k}) = (1/\sqrt{2})\begin{pmatrix} 1 & i\hat{k}\cdot\hat{L}_{12} & i\hat{k}\cdot\hat{L}_{13} \end{pmatrix} e^{i\vec{k}\cdot\vec{r}}, \quad \Psi_-^T(\vec{k}) = (1/\sqrt{2})\begin{pmatrix} 1 & -i\hat{k}\cdot\hat{L}_{12} & -i\hat{k}\cdot\hat{L}_{13} \end{pmatrix} e^{i\vec{k}\cdot\vec{r}}$$

and $\Psi_0^T(\vec{k}) = (1/\sqrt{2})(0 \quad -i\hat{k}\cdot\hat{L}_{13} \quad i\hat{k}\cdot\hat{L}_{12})e^{i\vec{k}\cdot\vec{r}}$ for $\Delta\omega > 0$, $\Delta\omega < 0$ and $\Delta\omega = 0$, respectively. From these Bloch states, we calculate the Berry phase [25] and find $\Gamma_\pm = i\oint \langle \Psi_\pm(\vec{k}) | \nabla_{\vec{k}} | \Psi_\pm(\vec{k}) \rangle \cdot d\vec{k} = 0$ [*See Section S1 in the Supplementary Material*], which is consistent with the fact that the Eq. (11) cannot be put into the Dirac Hamiltonian form. It is important to note that although the dispersion here also has two cones toughing at one point as in the case of graphene, the existence of a third degree of freedom changes the physics. Some papers in the literatures (see e.g. Ref. [13]) used the Dirac equation to describe the transport properties of these systems and such assumption can lead to erroneous conclusion in the presence of disorder as the zero Berry phase implies that a Dirac-like cone exhibits normal localization behavior rather than anti-localization.

Point *B* in Fig. 1(a) represents a Dirac point where two bands meet at the *K* point. These two bands are always degenerate at the *K* point of a triangular/honeycomb lattice, independent of the radii of the scatterers. This is a result of crystal symmetery and called the "deterministic degeneracy". The two degenerate Bloch states $\psi_{1K}$ and $\psi_{2K}$ are shown in Figs. 3(a) and 3(b), respectively. The reduced Hamiltonian has the following form:

$$H = \Delta \vec{k} \cdot \begin{pmatrix} \vec{p}_{11} & \mathrm{Re}\,\vec{p}_{12} + i\,\mathrm{Im}\,\vec{p}_{12} \\ \mathrm{Re}\,\vec{p}_{12} - i\,\mathrm{Im}\,\vec{p}_{12} & -\vec{p}_{11} \end{pmatrix}, \qquad (12)$$

where $\vec{p}_{11}$, $\mathrm{Re}\,\vec{p}_{12}$, and $\mathrm{Im}\,\vec{p}_{12}$ are three real vectors. $H$ can be rewritten as $H = (\Delta k)_x (\vec{\alpha}_x \cdot \vec{\sigma}) + (\Delta k)_y (\vec{\alpha}_y \cdot \vec{\sigma})$, where $\vec{\sigma} = (\sigma_1, \sigma_2, \sigma_3)$ are the Pauli matrices and $\vec{\alpha}_x = ((\mathrm{Re}\,\vec{p}_{12})_x, -(\mathrm{Im}\,\vec{p}_{12})_x, (\vec{p}_{11})_x)$ and $\vec{\alpha}_y = ((\mathrm{Re}\,\vec{p}_{12})_y, -(\mathrm{Im}\,\vec{p}_{12})_y, (\vec{p}_{11})_y)$ are two orthogonal real vectors having the same length, i.e., $\vec{\alpha}_x \cdot \vec{\alpha}_y = 0$ and $|\vec{\alpha}_x| = |\vec{\alpha}_y| = s/a$. After some unitary transformation in the pseudo-spin space [*See Section S2 in the Supplementary Material*], the Hamiltonian can be casted into the Dirac form, $H = \frac{s}{a}\left[(\Delta k)_x \sigma_x + (\Delta k)_y \sigma_y\right]$, where $\sigma_{x,y} = \frac{\vec{\alpha}_{x,y}}{s/a} \cdot \vec{\sigma}$ are the new Pauli matrices in the transformed pseudo-spin space. It is easy to show that the slopes of the linear dispersions are $\frac{\Delta\omega}{\Delta k}=\pm\frac{sc_1^2}{2\omega_0 a}$. The numerical results are plotted in Fig. 3(c) in red curves, which again agree excellently with the band structure calculations.

From Eq. (12), we obtain the following Bloch states near point B:
$$\Psi_\pm^T(\Delta\vec{k})=(1/\sqrt{2})\left(\sqrt{1\mp\Delta\hat{k}\cdot\vec{M}_{11}} \quad \frac{\mp\Delta\hat{k}\cdot\vec{M}_{12}^*}{\sqrt{1\mp\Delta\hat{k}\cdot\vec{M}_{11}}}\right)e^{i\Delta\vec{k}\cdot\vec{r}}, \text{ where } \vec{M}_{11}\equiv\frac{\vec{p}_{11}}{s/a} \text{ and}$$

$\vec{M}_{12}^*\equiv\frac{\vec{p}_{12}^*}{s/a}$ are two dimensionless vectors. From which we calculate the Berry phase

and find $\Gamma_\pm=-\frac{\sin(\alpha-\beta)}{2}\cdot\frac{|\text{Re}\,\vec{p}_{12}|}{s/a}\cdot\frac{|\text{Im}\,\vec{p}_{12}|}{s/a}\int_0^{2\pi}\frac{1}{1\mp\cos\varphi\cdot\frac{|\vec{p}_{11}|}{s/a}}d\varphi=-\pi$, where $\varphi$, $\alpha$

and $\beta$ are the angles between the vector $\vec{p}_{11}$ and the vectors $\Delta\vec{k}$, $\text{Re}\,\vec{p}_{12}$ and $\text{Im}\,\vec{p}_{12}$, respectively [*See Section S2 in the Supplementary Material*]. This result is consistent with the fact that the Eq. (12) can be mapped into the Dirac Hamiltonian.

To conclude, we have developed a first-principles theory to study the origin of Dirac/Dirac-like cone dispersions in phononic/photonic crystals. The theory can predict accurately the slopes of linear dispersions of a Dirac/Dirac-like cone at any symmetric point in a Brillonin zone, independent of frequency and lattice structure. The reduced Hamiltonian constructed from the theory shows that only the Dirac cones created by double degenerate Bloch states can be mapped into the massless Dirac Hamiltonian and carry a Berry phase. The absence of Berry phase in Dirac-like cones implies normal localization behavior when disorder is introduced.

**Acknowledgements**

We thank Profs. Bradley A. Foreman, K. T. Law and P. Sheng for stimulating discussions. This work was supported by National Natural Science Foundation of China (Grant No. 10804086), the Ph.D. Programs Foundation of Ministry of Education of China (Grant No. 200804861018), the Fundamental Research Funds for the Central Universities (Grant No. 2012ZZ0077), KAUST Start-up Package, and Hong Kong RGC (Grant No. 600311).

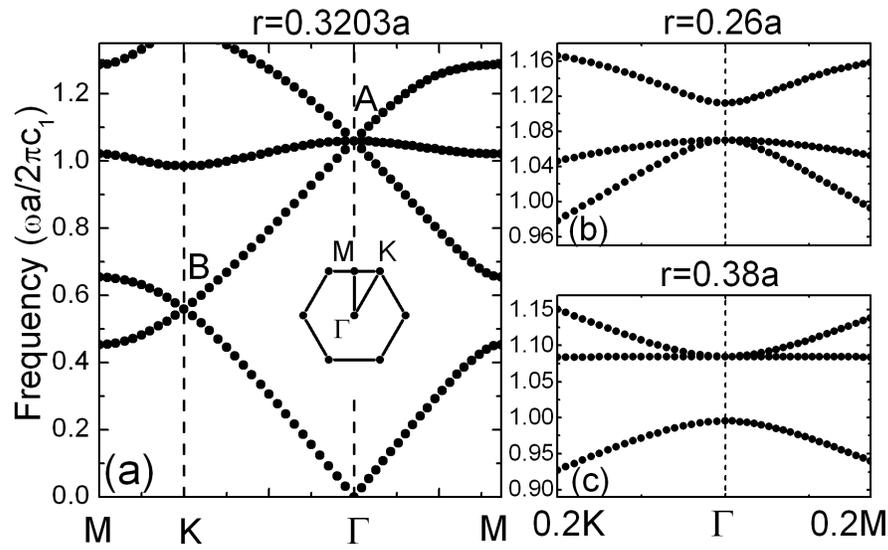

Figure 1.

(a) Band structures of a 2D phononic crystal consisting of a triangular lattice of Fe rods embedded in water host, with rod's radius *r=0.3203a*, where *a* is the lattice constant. Point A is a Dirac-like point while point B is a Dirac point. The inset shows the Brillouin Zone.

(b) and (c) are the band structures obtained from different rod radii, with r=0.26a in (b), and r=0.38a in (c), respectively.

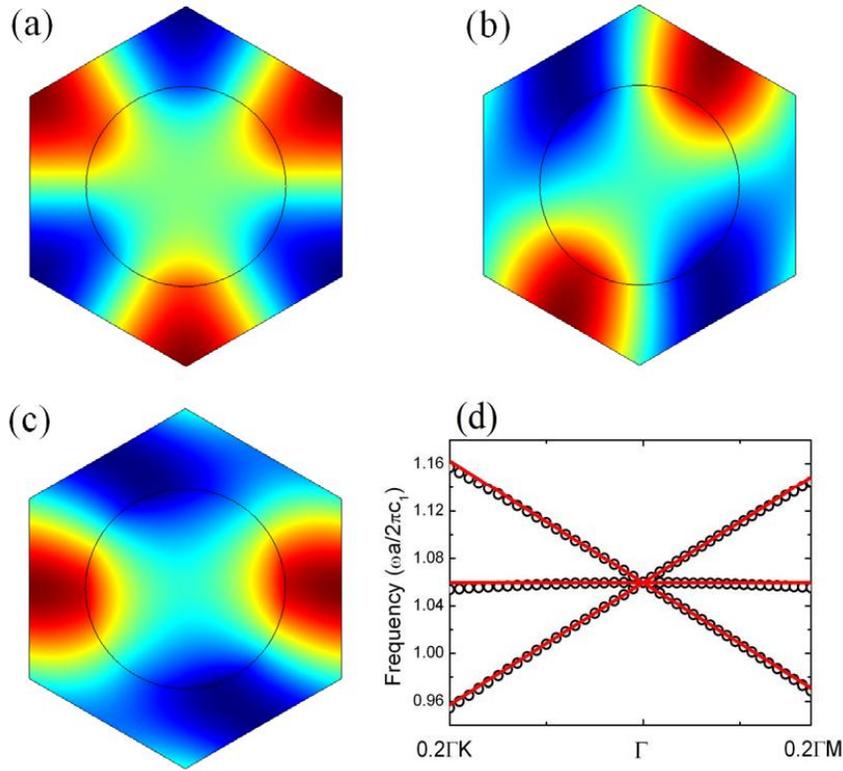

Figure 2.

(a), (b), and (c) show the pressure field distributions of three degenerate Bloch states, $\psi_{1\Gamma}$, $\psi_{2\Gamma}$, and $\psi_{3\Gamma}$, respectively, at point A as indicated in Fig. 1. Dark red and dark blue denote the positive and negative maxima, respectively, which imply that the wave energy is mainly concentrated in the host medium.

(d) Black circles show the band structure near point A obtained by finite-element calculations. Red curves show the results predicted by Eqs. (8) and (11).

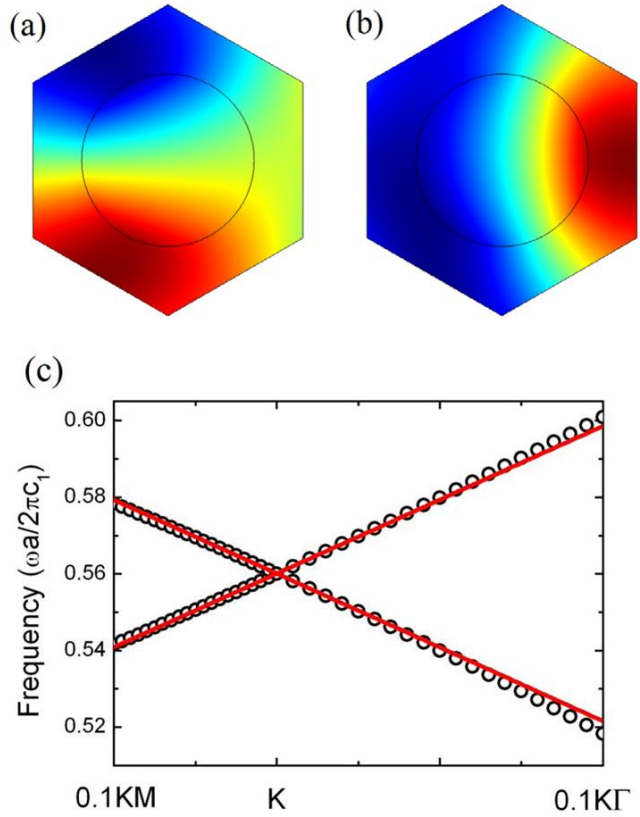

Figure 3.

(a) and (b) show the pressure field distributions of two degenerate Bloch states, $\psi_{1K}$ and $\psi_{2K}$, respectively, at point B as indicated in Fig. 1. Dark red and dark blue denote the positive and negative maxima, respectively.

(c) Black circles show the band structures near point B obtained by finite-element calculations. Red curves show the results predicted by Eqs. (8) and (12).